\RequirePackage{fix-cm}
\documentclass{svjour3}                     % onecolumn (standard format)
\smartqed  % flush right qed marks, e.g. at end of proof
\usepackage{graphicx}
\begin{document}

\title{Speeding up lower bound estimation in powerlaw distributions%\thanks{Grants or other notes
}


\author{Alessandro Bessi}


\institute{Alessandro Bessi \at
              IUSS Institute for Advanced Study \\
              Palazzo del Broletto, Piazza della Vittoria, 5\\
              27100 Pavia (Italy)\\
              \email{alessandro.bessi@iusspavia.it}           %  \\
}

\date{Received: date / Accepted: date}

\maketitle

\begin{abstract}
The traditional lower bound estimation method for powerlaw distributions based on the Kolmogorov-Smirnov distance proved to perform better than other competing methods. However, if applied to very large collections of data, such a method can be computationally demanding. In this paper, we propose two alternative methods with the aim to reduce the time required  by the estimation procedure. We apply the traditional method and the two proposed methods to large collections of data ($N = 500,000$) with varying values of the true lower bound. Both the proposed methods yield a significantly better performance and accuracy than the traditional method. 
\end{abstract}

\section{Introduction}
\label{intro}
Over the last few decades powerlaw distributions have attracted particular attention for their mathematical properties and appearances in a wide variety of scientific contexts, from physical and biological sciences to social and man-made phenomena \cite{beggs2003,bell2012,clauset2007,duijn2014,michel2011,yu2008}. Differently from Normally distributed data, empirical quantities that follow a powerlaw distribution do not cluster around an average value, and thus can not be characterized through the mean and standard deviation. Nevertheless, the fact that some scientific observations can not be characterized as simply as other measurements is often a sign of complex underlying processes that deserve further study \cite{clauset2009}. A complete introduction to powerlaw distributions along with a statistical framework for discerning and quantifying powerlaw behavior in empirical data can be found in \cite{clauset2009}, whereas extensive discussions can be found in \cite{mitzenmacher2004,newman2005,sornette2006}, and references therein. Recent advances related to powerlaw fitting and statistical hypothesis testing can be found in \cite{gillespie2014b,bessi2015b}.

Formally, a quantity $x$ follows a powerlaw distribution if its probability distribution is defined as
$$ p(x) \propto x^{-\alpha},$$

where $x > 0$ and $\alpha > 1$ is called the \emph{scaling parameter} of the distribution. Fitting these kind of heavy-tailed distributions requires care, since only few empirical phenomena show such a probability distribution for all values of $x$. Indeed, more often only values greater than some minimum value $x_{min}$, i.e. the so called \emph{lower bound}, follow a powerlaw distribution.

The traditional lower bound estimation method introduced in \cite{clauset2009} is based on the computation of the Kolmogorov-Smirnov distances between the empirical and the theoretical cumulative distribution functions defined for values $x \geq x_{min}$ when $x$ is discrete ($x > x_{min}$ when $x$ is continuous). Once the Kolmogorov-Smirnov distances have been computed for all the eligible values of $x_{min}$, the $x_{min}$ associated with the smallest distance is chosen as lower bound of the distribution. However, if applied to very large collections of data -- e.g. the distribution of the number of views received by YouTube videos -- such a method can be computationally demanding, and bootstrap techniques to address the uncertainty in the estimates and average over multiple estimations become unfeasible. 

In this paper, we propose two alternative methods with the aim to reduce the time required  by the traditional estimation procedure. In particular, the first proposed method starts to compute the traditional Kolmogorov-Smirnov distance from a guess on the true value of the lower bound, and stops the procedure once a minimum is reached. The second proposed method is thought for the discrete case, where the computation of theoretical cumulative distribution functions involves the calculation of Hurwitz zeta functions, which could be computationally binding. Such a method uses the above-mentioned conditions to reduce the number of computations, and substitutes the cumulative distribution functions of the traditional Kolmogorov-Smirnov distance with the corresponding probability mass functions, i.e. it is based on the comparison between empirical and theoretical probabilities for each $x \geq x_{min}$. 

This manuscript is organized as follows. In Section 2 we provide some basic definitions about continuous and discrete powerlaw distributions. In Section 3 we first discuss the traditional estimation method, and then we introduce two alternative methods which can speed up the estimation procedure. In Section 4 we apply the three methods to large collections of data ($N = 500,000$) with varying values of the true lower bound, showing that both our proposed methods yield a significantly better performance and accuracy than the traditional method. Section 5 is left for some concluding remarks.

\section{Definitions}
\label{sec:1}
Let $x$ represents a quantity whose distribution we are interested in. The probability distribution when $x$ is continuous is defined as

$$ p(x)dx = Pr(x \leq X < x + dx) = \frac{\alpha - 1}{x_{min}}\left( \frac{x}{x_{min}}\right)^{-\alpha}, $$

whereas in the discrete case, when $x$ can assume only positive integers, the probability distribution is defined as

$$ p(x) = Pr(X = x) = \frac{x^{-\alpha}}{\zeta(\alpha,x_{min})}, $$

where $x_{min} > 0$ is the lower bound, $\alpha > 1$ is the scaling parameter, and
$$\zeta(\alpha,x_{min}) = \sum_{n=0}^{\infty}(n + x_{min})^{-\alpha}$$ 

is the Hurwitz zeta function. Furthermore, the complementary cumulative distribution function in the continuous case is defined as

$$ P(x) = 1 - Pr(X \leq x) = Pr(X > x) = \left( \frac{x}{x_{min}} \right)^{-\alpha + 1},$$

whereas in the discrete case is defined as

$$ P(x) = 1 - Pr(X \leq x) = Pr(X > x) = \frac{\zeta(\alpha,x)}{\zeta(\alpha,x_{min})}.$$

The complementary cumulative distribution function is often preferred to the cumulative distribution function since it allows to show powerlaw distributions in doubly logarithmic axes, and thus emphasize their upper tail behavior.

\section{Lower bound estimation}
\label{sec:2}

\subsection{Traditional Method}
The traditional method to estimate the lower bound of a powerlaw distribution has been introduced in \cite{clauset2009}. Such a method is based on the Kolmogorov-Smirnov distance, which is defined as
$$ D = \max\bigl(\,\,| E - T |\,\,\bigr), $$

where $E$ is the empirical cumulative distribution function, and $T$ is the theoretical cumulative distribution function of the fitted powerlaw distribution for values $x \geq x_{min}$ when $x$ is discrete ($x > x_{min}$ when $x$ is continuous, and hereafter we refer only to the discrete case for the sake of brevity). Once the Kolmogorov-Smirnov distance has been computed for all the possible values $x_{min}$, the $x_{min}$ associated with the smallest value of $D$ is chosen as the lower bound of the powerlaw distribution.

In \cite{clauset2009} it has been proved that the estimation method based on the Kolmogorov-Smirnov distance outperforms alternative methods based on the BIC (Bayesian Information Criterion) and the Anderson-Darling statistics. Nevertheless, when dealing with big data such a method can be computationally demanding for two main reasons:
\begin{enumerate}
	\item the algorithm needs to compute the Kolmogorov-Smirnov distance for each possible $x_{min}$; 
	\item in the discrete case, the computation of the theoretical cumulative distribution function of the fitted powerlaw distributions involves the calculation of Hurwitz zeta functions, which can be computationally binding when dealing with large data collections.
\end{enumerate}

\subsection{Proposed Methods}
\begin{center}
	\begin{figure}[h]
		\centering
		\includegraphics[width = \textwidth ]{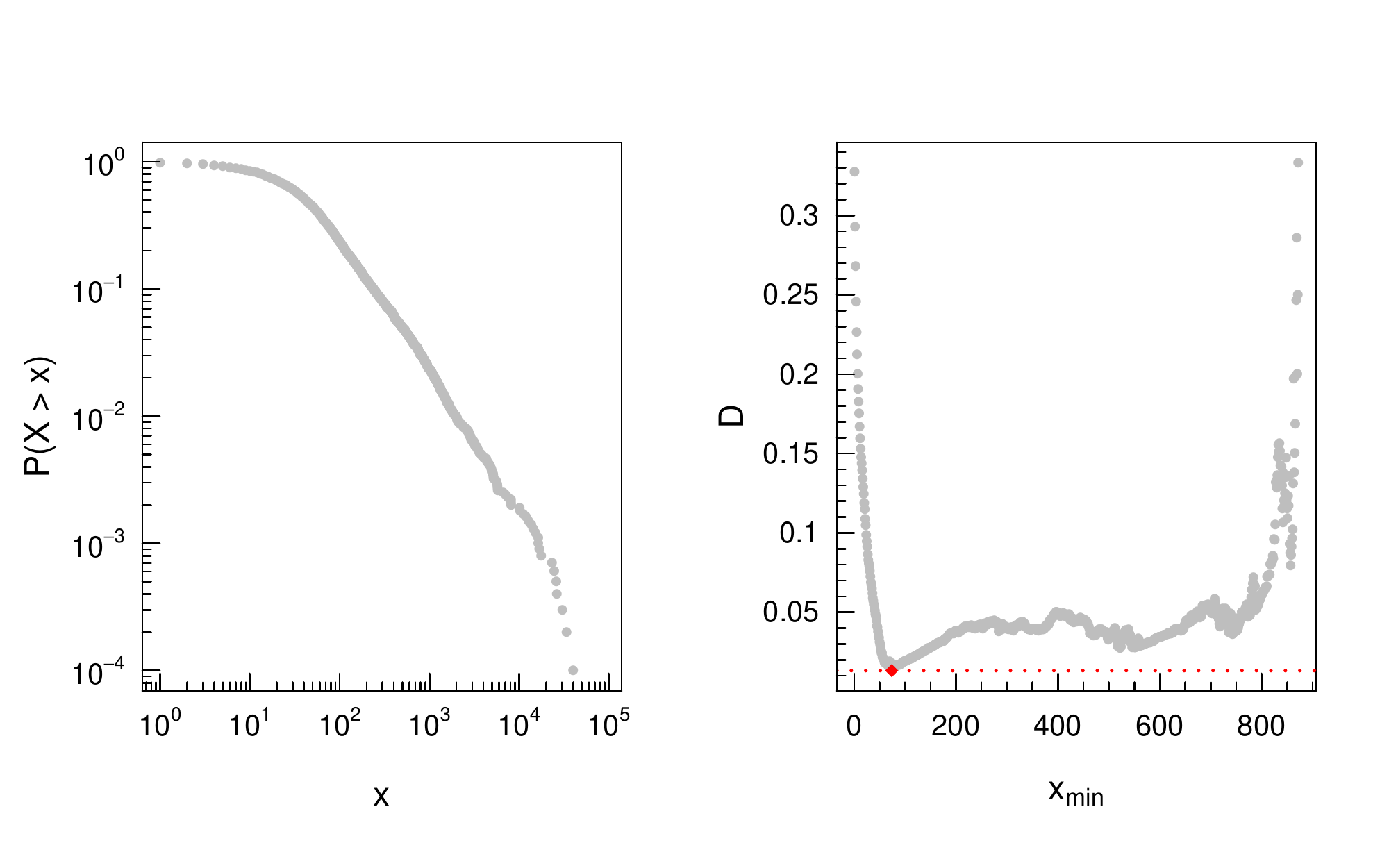}
		\caption{\textbf{Exploratory Analysis and Kolmogorov-Smirnov distances.} On the left panel we show the complementary cumulative distribution function of a random generated distribution with powerlaw tail ($N = 10^{4}$, $\alpha = 2$, $x_{min} = 100$). Notice that a quick graphical analysis is sufficient to rule out some eligible values of $x_{min}$. When dealing with big data and large maximal values, we could rule out hundreds of eligible values, thus reducing the required time to estimate the lower bound. On the right panel we show the values of the Kolmogorov-Smirnov statistics in correspondence to subsequent values of $x_{min}$.}
		\label{fig:example}
	\end{figure}
\end{center}

In this paper, we aim at introducing two lower bound estimation methods in order to tackle the above-mentioned drawbacks of the traditional method. We start from two simple observations. First of all, often there is no need to compute the Kolmogorov-Smirnov distance for all the eligible values of $x_{min}$, since a graphical exploratory analysis is usually sufficient to rule out a substantial range of values. On the left panel of Figure \ref{fig:example}, we show the complementary cumulative distribution function of a random generated distribution with powerlaw tail. Notice that a quick look at the plot is sufficient to rule out some eligible values for the lower bound. When dealing with big data and large maximal values, we could rule out hundreds of possible values, thus reducing the required time to estimate the lower bound. Moreover, we know by definition that the Kolmogorov-Smirnov statistics computed for all the possible values $x_{min}$ has a global minimum in correspondence to the true lower bound. On the right panel of Figure \ref{fig:example}, we show the values of the Kolmogorov-Smirnov distance in correspondence to subsequent values of $x_{min}$.

Taking into account these observations, we propose a first estimation method which starts computing the Kolmogorov-Smirnov distances from the eligible value of $x_{min}$ that is closest to
$$g - g\frac{(100 - c)}{100} \in \bigl[min(x),max(x)\bigr), $$

where $g$ is a guess on the true value of the lower bound, and $c \in [1,100]$ is the confidence in such a guess. The computation of the Kolmogorov-Smirnov distances stops when all the differences between the last $k$ distances are positive -- i.e. when a minimum is reached. The key ideas are two:
\begin{enumerate}
	\item since a quick graphical exploratory analysis is often sufficient to rule out a large amount of eligible values, we can start computing the Kolmogorov-Smirnov distances from the value we think it is the true lower bound;
	\item if our guess is close to the true lower bound, the first minimum of the Kolmogorov-Smirnov statistics we meet is the global minimum associated with the true lower bound, and hence we can stop the computation once a minimum is reached.
\end{enumerate}

Moreover, since in the discrete case the computation of the theoretical cumulative distributions involves the calculation of Hurwitz zeta functions, we propose a second estimation method that further modify the traditional method by substituting the empirical and theoretical cumulative distribution functions of the Kolmogorov-Smirnov distance with the corresponding probability mass functions, which are generally faster to compute. More formally, in the second proposed method the distance to be computed is defined as
$$ D_{pmf} = \max\bigl(\,\,| E_{pmf} - T_{pmf} |\,\,\bigr), $$

where the subscripts indicate that $E$ and $T$ are, respectively, the empirical probability mass functions, and the theoretical probability mass function of the fitted powerlaw distribution for values $x \geq x_{min}$.

\section{Results and Discussion}
\label{sec:3}
In this section, we illustrate and discuss the results of a simulation\footnote{All the simulations have been performed on a machine with Ubuntu 14.10, R 3.1.1, Intel quad-core 4 GHz CPU, 32 Gb RAM.} comparing the two proposed methods with the traditional estimation method. We refer to the traditional method as \texttt{estimate\_xmin} -- the name of the lower bound estimation function provided by the \textsc{R} package \texttt{poweRlaw} \cite{gillespie2014} -- and to our proposed methods as
\begin{enumerate}
	\item \texttt{getXmin}: the first proposed method still based on the classical traditional Kolmogorov-Smirnov distance;
	\item \texttt{getXmin2}: the second proposed method based on distances between empirical and theoretical probability mass functions.
\end{enumerate}

Both \texttt{getXmin} and \texttt{getXmin2} are implemented for discrete powerlaw distributions on the \textsc{R} package \texttt{staTools} \cite{bessi2015}, which is currently available on \textsc{CRAN}.

In order to test the three different methods, we generate synthetic data and examine both the accuracy and the performance  in the estimation of the true lower bound. We use data drawn from a distribution with the form
\begin{equation}
p(x)=\left\{
\begin{array}{l l}     
	C(x/x_{min})^{-\alpha} & \quad x \geq x_{min}\\
	Ce^{-\alpha(x/x_{min} - 1)} & \quad x< x_{min}
\end{array}\right.
\label{eq:distr}
\end{equation}

where $\alpha = 3$ and $C$ is a normalization constant. We apply the three estimation methods to large ($N = 500,000$) collections of data drawn from Eq. \ref{eq:distr} with true values of $x_{min}$ varying in $50,100,\dots,500$. 

In our simulation, we set $g$ equal to the true lower bound with a $90\%$ confidence, $c = 90$ -- e.g. when the true lower bound is $500$, our proposed methods start to compute the corresponding statistics from the possible value of $x_{min}$ that is closest to $500 - 500 \times (100 - 90)/100 = 450$, which is a feasible practice\footnote{The \texttt{inspect} function provided by the \textsc{R} package \texttt{staTools} allows the user to quickly visualize the powerlaw fit for different values of $x_{min}$, thus assisting the user in making a good guess on the true value of the lower bound.} and thus a reasonable assumption. Moreover, both \texttt{getXmin} and \texttt{getXmin2} stop to compute the corresponding statistics once a first minimum is reached, i.e. when all the differences between the last $k = 5$ computed distances are positive.

\begin{center}
	\begin{figure}[h]
		\centering
		\includegraphics[width = 0.5 \textwidth ]{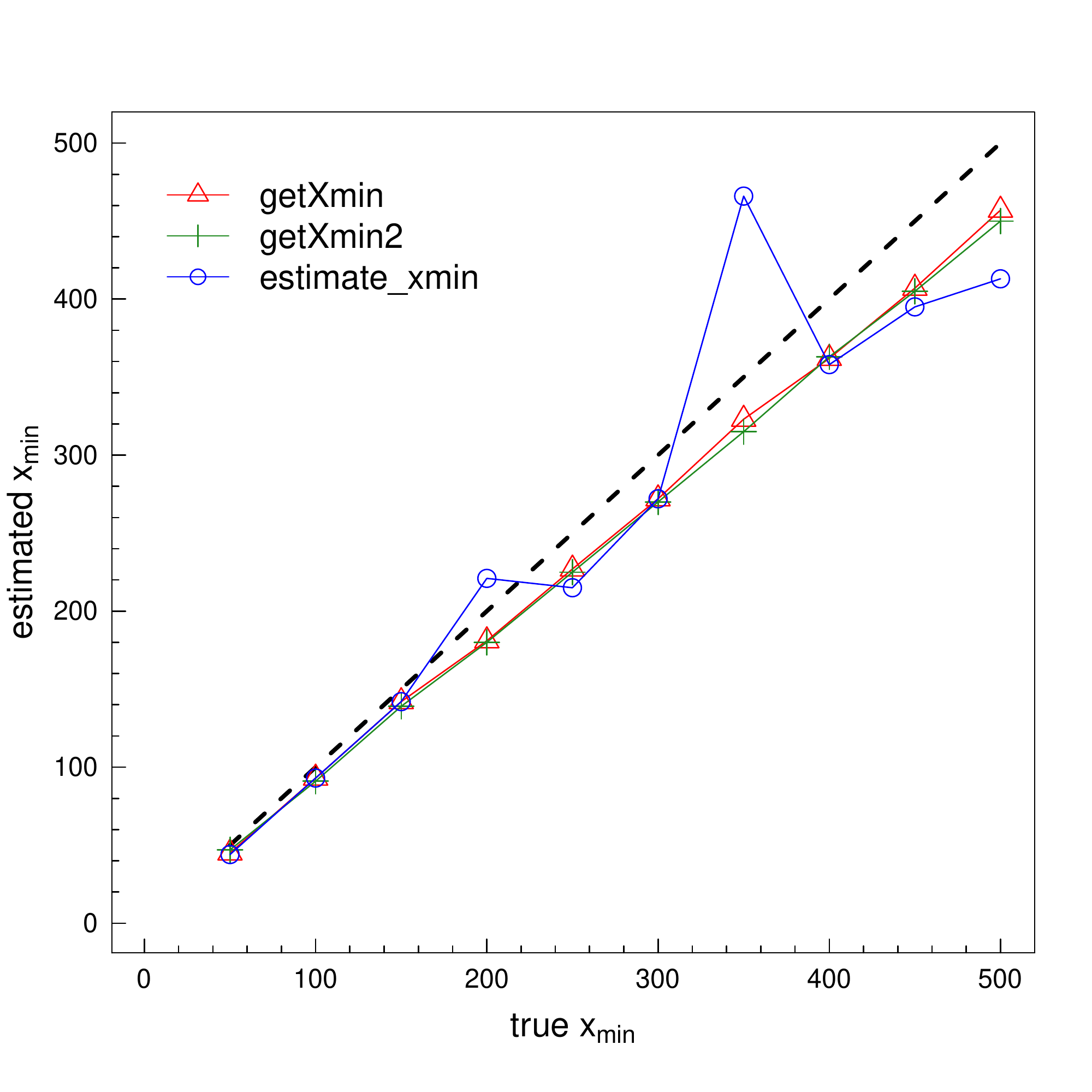}
		
		\caption{\textbf{Simulations.} Both \texttt{getXmin} and \texttt{getXmin2} ($g = true\,\,x_{min}$, $c = 90$, $k = 5$) outperform the traditional estimation method.}
		\label{fig:simulation}
	\end{figure}
\end{center}

Figure \ref{fig:simulation} shows the estimated value of $x_{min}$ as a function of the true lower bound, indicating that both \texttt{getXmin} and \texttt{getXmin2} outperform the traditional estimation method. Table \ref{tb:accuracy} summarizes the accuracy of the three methods through mean squared errors (MSEs), root mean squared errors (RMSEs), and mean absolute errors (MAEs), confirming that both the proposed methods yield a better accuracy than the traditional method.

\begin{center}
	\begin{table}[ht]% Try here, and then top
		\centering
		\begin{tabular}{rccc}
			
			& MSE & RMSE & MAE \\
			\hline
			\texttt{getXmin}        & \textbf{768.3} & \textbf{27.72} & \textbf{24.1} \\
			\texttt{getXmin2}       & \textbf{925.5} & \textbf{30.42} & \textbf{26.5} \\
			\texttt{estimate\_xmin} & 2841.3 & 53.30 & 40.5	\\
			\hline
			
		\end{tabular}
		\caption{\textbf{Estimation accuracy.} Mean squared errors, root mean squared errors, and mean absolute errors summarizing the accuracy of the lower bound estimates obtained by means of three different methods. Both the proposed methods yield a better accuracy than the traditional method.}
		\label{tb:accuracy}
	\end{table}
\end{center}

\begin{center}
	\begin{figure}[h]
		\centering
		\includegraphics[width = 0.5 \textwidth]{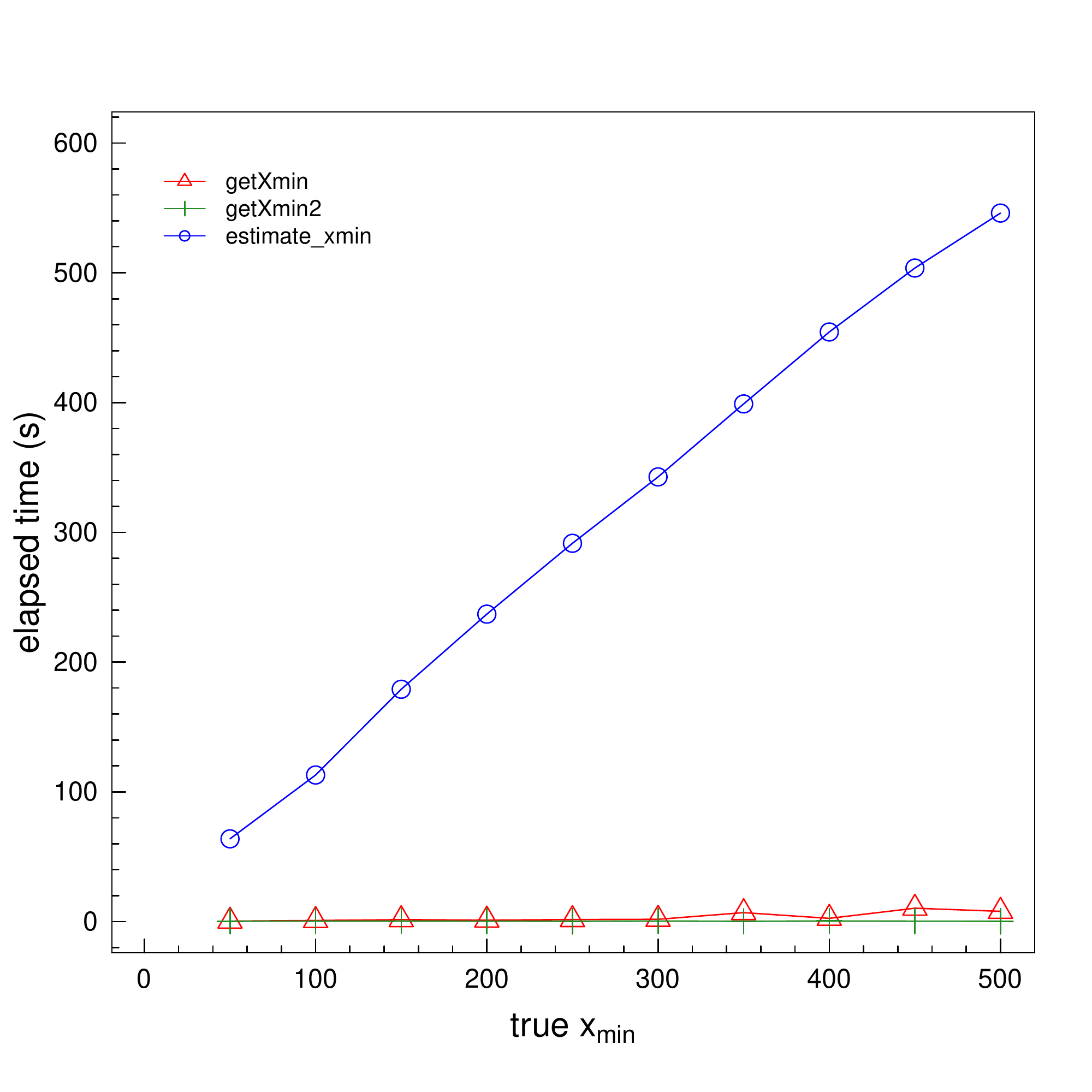}

		\caption{\textbf{Performance.} Time demanded by the different methods to estimate the lower bound. Both our proposed methods yield a better performance than the traditional estimation method.}
				\label{fig:performance}
	\end{figure}
\end{center}

Figure \ref{fig:performance} illustrates the time demanded by the different estimation methods, indicating that our proposed methods yield a better performance than the traditional estimation method. 

\section{Conclusions}
The traditional lower bound estimation method for powerlaw distributions proved to outperform competing methods based on BIC and Anderson-Darling statistics. However, if applied to very large collections of data, such a method can be computationally demanding, and bootstrap techniques to address the uncertainty in the estimates and average over multiple estimations become unfeasible. In this paper, we propose two alternative methods with the aim to reduce the time required  by the estimation procedure. In particular, the first proposed method starts to compute the traditional Kolmogorov-Smirnov distances from a guess on the true value of the lower bound, and stops the procedure once a minimum is reached. The second proposed method uses the above-mentioned conditions to reduce the number of computations, and substitutes the cumulative distribution functions of the traditional Kolmogorov-Smirnov statistics with the corresponding probability mass functions. We apply the three methods to large collections of data ($N = 500,000$) with varying values of the true lower bound. Both the proposed methods yield a significantly better performance and accuracy than the traditional method. 

\section*{Acknowledgements}
We would like to thank Colin S. Gillespie for our helpful discussion.


\end{document}